\documentstyle[12pt]{article}
\def\be {\begin{equation}}
\def\ee {\end{equation}}
\def\ba {\begin{eqnarray}}
\def\ea {\end{eqnarray}}
\def\nn {\nonumber}
\def\bi {\begin{itemize}}
\def\ei {\end{itemize}}

\def\bea{\begin{eqnarray}}
\def\nn{\nonumber}
\def\eea{\end{eqnarray}}
\begin{document}
\def\bea{\begin{eqnarray}}
\def\eea{\end{eqnarray}}
\title{\bf {Casimir effect for scalar fields with Robin boundary conditions
in Schwarzschild background  }}
 \author{M.R. Setare  \footnote{E-mail: rezakord@ipm.ir}
  \\{Physics Dept. Inst. for Studies in Theo. Physics and
Mathematics(IPM)}\\
{P. O. Box 19395-5531, Tehran, IRAN }}

\maketitle
\begin{abstract}
The stress tensor of a massless scalar field satisfying Robin
boundary conditions on two one-dimensional wall in two-dimensional
Schwarzschild background is calculated. We show that vacuum
expectation value of stress tensor can be obtained explicitly by
Casimir effect, trace anomaly and Hawking radiation.
 \end{abstract}
\newpage

 \section{Introduction}
 The Casimir effect is one of the most interesting manifestations
  of non-trivial properties of the vacuum state in quantum field
  theory \cite{{mueller},{trunov}}. Since its first prediction by
  Casimir in 1948 \cite{Casimir}, this effect has been investigated for
  various cases of boundary geometries and various types of
  fields \cite {Remeo}-\cite{Birrell}. The Casimir effect can be viewed as a
   polarization of  vacuum by boundary conditions. Another type of vacuum
  polarization arises in the case of an external gravitational
  field. In this paper we study a situation when both types of
  sources for the polarization are present. There are several
  methods to calculate Casimir energy. For instance,  we can mention mode summation,
  Green's function method \cite{mueller}, heat kernel method \cite{klaus}along with appropriate
   regularization schemes such as point separation \cite{chr},\cite{adler}
  dimensional regularization \cite{deser}, zeta function regularization
  \cite{{haw},{blu},{Remeo},{Elizalde}}. Recently a general new methods to compute renormalized
   one--loop quantum energies and energy densities are given in
   \cite{{gram1},{gram2}} (see also \cite{eliz2}).\\
   It has been shown \cite{{Nugayev1},{Nugayev2}} that particle creation
   by a black hole in four dimensions is a consequence of the Casimir
effect for a spherical shell. It has been shown that just the
existence of the horizon and of the barrier in the effective
potential are sufficient to compel the black hole to emit
black-body radiation with a temperature that exactly coincides
with the standard result for Hawking radiation. In
\cite{Nugayev2}, the results for the accelerated mirror have been
used to prove the
above statement.\\
The renormalized vacuum expectation value of the stress tensor of
the scalar field in the Schwarzschild spacetime can be obtained by
using different regularization methods.( see Refs. \cite
{chris}-\cite{set2}). $<T_{\nu }^{\mu }>_{ren}$ is needed , for
instance, when we want to study back-reaction, i.e, the influence
that the matter field in a curved background assert on the
background geometry itself. This would be done by solving the
Einstein equations with the
expectation value of the energy-momentum tensor as source.\\
In this paper the Casimir energy for massless scalar field in two-
dimensional
  Schwarzschild black hole for two parallel plates
   with Robin boundary conditions is calculated. The Casimir energy is obtained
   by imposing general requirements. Calculation of the renormalized stress tensor
   for massless scalar field in two-dimensional Schwarzschild background
   has been done in \cite{{chris},{vass}, {Fab}}. The Casimir effect
   for a massless scalar filed under Dirichlet boundary condition in a two-dimensional
   Domain wall, Schwarzschild black hole, stringy black hole, Achucarro-Ortiz black hole,
   have been studied respectively in
   \cite{{set3},{set4},{set1},{elias1},{elias2}}. The Robin
   boundary condition includes the Dirichlet and Neumann boundary
   conditions as special cases. The Casimir effect for the general
   Robin boundary conditions on background of the Minkowski
   spacetime was investigated in Ref. \cite{sah1} for flat
   boundaries, here we use the results of this reference to
   generate vacuum energy-momentum tensor in our interesting
   background.
   Knowing the Casimir energy in flat space and the
   trace anomaly can help us to calculate renormalized stress tensor. In other
   situations, besides the two previous quantities, we use Hawking radiation,
   which also has a contribution in the stress tensor. Therefore, the
   renormalized stress tensor is not unique and depends on the vacuum under consideration.
   Our paper is organised as follows.
 In section 2 general properties of stress tensor are discussed. Then in section
 3, the vacuum expectation value of the stress tensor in two dimensions is obtained.
 Finally, in section 4, we conclude and summarize the results.

\section{General properties of stress tensor}
  In a semiclassical framework for yielding a sensible theory of back
  reaction, Wald \cite{wald} has developed an axiomatic approach.
  There one tries to obtain an expression for the renormalized
  $T_{\mu\nu}$ from the properties (axioms) which it must fulfill.
  The axioms for the renormalized energy momentum tensor are as
  follows.

  1-For off-diagonal elements, the standard result should be obtained.

  2-In Minkowski spacetime, the standard result should be obtained.

  3-Expectation values of energy-momentum are conserved.

  4-Causality holds .

  5-The energy-momentum tensor contains no local curvature tensor depending on
  derivatives of the metric higher than second order.

  Two prescriptions that satisfy the first four axioms can differ
  by at most a conserved local curvature term. Wald \cite{wald2},
   showed any prescription
  for renormalized $T_{\mu \nu}$ which is consistent with axioms 1-4 must yield the
  given trace up to the addition of the trace of conserved local curvature.
  It  must be noted  that trace anomalies in a stress tensor, i.e.
   the non-vanishing  $T^\mu _\mu$ for a conformally invariant field after
    renormalization  originate from some quantum behavior \cite{col}. In two-dimensional
  spacetime one can show that a trace-free stress tensor cannot be
  consistent with conservation and causality
  if particle creation occurs. A trace-free, conserved stress
  tensor in two dimensions must always remain zero if it is
  initially zero.
  One can show that the 'Davies-Fulling-Unruh' formula \cite{Davies}
  for the stress tensor of a scalar field which yields an anomalous trace,
  $T^{\mu} _\mu=\frac{R}{24\pi}$, is unique which is
  consistent with the above axioms.
  In four dimensions, just as in two dimensions, a trace-free stress tensor which
  agrees with the formal expression for the matrix elements between orthogonal
  states cannot be compatible with both conservation laws and causality .
  It must be noted that, as Wald showed \cite{wald2}, with Hadamard regularization
  in the massless case axiom (5) cannot be satisfied unless we
  introduce a new fundamental length scale for nature. Regarding all of these
  axioms, thus we are able to obtain an unambiguous
  prescription for calculating the stress tensor.

\section{Vacuum expectation values of  stress tensor and Scalar Casimir effect }

  Our background shows a Schwarzschild black hole with the following
  metric:
  \begin{equation}\label{four}
  d^{2}s=-(1-2\frac{m}{r})dt^2+(1-2\frac{m}{r})^{-1}dr^2+r^2(d\theta^2+\sin^2d\varphi^2).
       \end{equation}
 Now we reduce the dimension of spacetime  to two,
  \begin{equation}\label{two}
  d^{2}s=-(1-\frac{2m}{r})dt^2+(1-\frac{2m}{r})^{-1}dr^2.
  \end{equation}
  The metric (2) can be written in conformal form
  \begin{equation}\label{conf}
  d^{2}s=\Omega(r)(-dt^2+dr^{\ast 2}),
  \end{equation}
  with
  \begin{equation}\label{coor}
  \Omega(r)=1-\frac{2m}{r},\hspace{20mm}\frac{dr}{dr^{\ast}}=\Omega(r).
  \end{equation}
  From now on, our main goal is to determine  a general form
  of conserved energy-momentum tensor with, regard to the trace anomaly for
  the metric $(2)$. It is necessary we mention that our method
  here is available for any general two-dimensional spacetime as
  following metric easily (see also \cite{set1,sal},
  \cite{set3}-\cite{elias2})
 \begin{equation}\label{gen}
  d^{2}s=-g(r)dt^2+g(r)^{-1}dr^{2}.
  \end{equation}
  For the non-zero Christoffel symbols of the
  metric $(2)$, we have in $(t,r^{\ast})$ coordinates:
  \begin{equation}\label{cris}
  \Gamma^{r^{\ast}}_{tt}=\Gamma^{t}_{tr^{\ast}}=\Gamma ^{t}_{r^{\ast}r^{\ast}}=
  \Gamma^{r^{\ast}}_{r^{\ast}r^{\ast}}=\frac{1}{2}\frac{d\Omega(r)}{dr}.
  \end{equation}
   Then the conservation equation takes the following form:
  \begin{equation}\label{coneq1}
  \partial_{r^{\ast}}{T^{r^{\ast}}_{t}}+
  \Gamma^{t}_{tr^{\ast}}T^{r^{\ast}}_{t}-
  \Gamma^{r^{\ast}}_{tt}T^{t}_{r^{\ast}}=0,
  \end{equation}
  \begin{equation}\label{coneq2}
  \partial_{r^{\ast}}T^{r^{\ast}}_{r^{\ast}}+
  \Gamma^{t}_{tr^{\ast}}T^{r^{\ast}}_{r^{\ast}}-
  \Gamma ^{t}_{tr^{\ast}}T^{t}_{t}=0
  \end{equation}
  in which,
  \begin{equation}\label{emcom}
  T^{t}_{r^{\ast}}=-T^{r^{\ast}}_{t},         \hspace{2cm}
  T^{t}_{t}=T^{\alpha}_{\alpha}-T^{r^{\ast}}_{r^{\ast}}
  \end{equation}
  and $T^{\alpha}_{\alpha}$ is the anomalous trace in two dimension. Using
  equations (\ref{cris}),(\ref{emcom})we can rewrite
  Eq.(\ref{coneq1}) as following
\begin{equation}\label{coneq3}
    \frac{d}{dr^{\ast}}T^{r^{\ast}}_{t}+\frac{1}{2}\frac{d\Omega}{dr}T^{r^{\ast}}_{t}
    +\frac{1}{2}\frac{d\Omega}{dr}T^{r^{\ast}}_{t}=\frac{d}{dr}T^{r^{\ast}}_{t}
    \frac{dr^{\ast}}{dr}+\frac{d\Omega}{dr}T^{r^{\ast}}_{t}=0,
  \end{equation}
  now using Eq.(\ref{coor})we have
  \begin{equation}\label{coneq4}
    \frac{d}{dr}(\Omega(r)T^{r^{\ast}}_{t})=0.
  \end{equation}
  Similarly  for Eq.(\ref{coneq2}) we have
\begin{equation}\label{coneq5}
  \frac{d}{dr^{\ast}}T^{r^{\ast}}_{r^{\ast}}+\frac{1}{2}\frac{d\Omega}{dr}T^{r^{\ast}}_{r^{\ast}}
  -\frac{1}{2}\frac{d\Omega}{dr}(T^{\alpha}_{\alpha}-T^{r^{\ast}}_{r^{\ast}})=0,
  \end{equation}
 again using Eq.(\ref{coor})we have
  \begin{equation}\label{coneq6}
  \frac{d}{dr}(\Omega(r)T^{r^{\ast}_{r^{\ast}}})=
  \frac{1}{2}(\frac{d\Omega(r)}{dr})T^{\alpha}_{\alpha}.
  \end{equation}
  Then Eq.(\ref{coneq4}) leads to
  \begin{equation}\label{alfa}
  T^{r^{\ast}}_{t}=\alpha \Omega^{-1}(r),
  \end{equation}
  where $\alpha$ is a constant of integration. The solution of Eq.(\ref{coneq6}) may be
  written in the following form:
  \begin{equation}\label{beta}
  T^{r^{\ast}}_{r^{\ast}}(r)=(H(r)+\beta)\Omega^{-1}(r),
  \end{equation}
  where
  \begin{equation}\label{heq}
  H(r)=1/2
  \int^{r}_{l}T^{\alpha}_{\alpha}(r')\frac{d}{dr'}\Omega(r')dr',
  \end{equation}
  with $l$ being an arbitrary scale of length and considering
  \begin{equation}\label{trac}
  T^{\alpha}_{\alpha}=\frac{R}{24\pi}=\frac{m}{6 \pi r^{3}}
  \end{equation}
  the function $H(r)$ produces the non-local contribution of the trace
  $T^{\alpha}_{\alpha}(x)$ to the energy-momentum tensor.
  Finding $l$ depends on the metric. For the metric (2) we choose \cite{Birrell}
  \begin{equation}\label{scal}
  l\approx r_{b}=2m,
  \end{equation}
  so we reach
  \begin{equation}\label{heq2}
  H(r) =\frac{m^{2}}{24\pi}(\frac{1}{16m^{4}}-\frac{1}{r^{4}}).
  \end{equation}
  Using Eqs. (\ref{emcom},\ref{alfa})and (\ref{beta}) it can be shown that the energy-momentum
  tensor takes the following form in
   $(t,r^{\ast})$ coordinates. So we have the most general form of
   stress tensor field in our background of interest:
   \begin{equation}
   T^\mu\   _\nu(r)=\left(\begin{array}{cc}
   T^{\alpha}_{\alpha}-\Omega(r)^{-1}H(r) & 0 \\
   0 &\Omega(r)^{-1}H(r) \
   \end{array}\right)+\Omega^{-1}\left(\begin{array}{cc}
   -\beta&-\alpha  \\
   \alpha & \beta \
   \end{array}\right).\label{emteq}
   \end{equation}
  Now we are going to obtain two constants  $\alpha$
     and $\beta$ by imposing the second axiom of the
   renormalization scheme. We consider two one-dimensional walls which are placed
   at point $r_1$ and $r_2$ in our interest background. The massless scalar field
   whose energy-momentum tensor we try to evaluate satisfies the Robin boundary
   conditions on the walls.
    At first we review the casimir effect
   for massless scalar filed under Robin boundary condition on
   plates in Minkowski spacetime briefly (see \cite{sah1}),
   We will assume that the field satisfies the mixed boundary
condition
\begin{equation}
\left( a_{j}+b_{j}n^{\mu }\nabla _{\mu }\right) \varphi
(x)=0,\quad r=r_{j},\quad j=1,2  \label{boundcond}
\end{equation}
on the plate $r=r_{1}$ and $r=r_{2}$, $r_{1}<r_{2}$, $n^{\mu }$ is
the normal to these surfaces, $n_{\mu }n^{\mu }=-1$, and $a_j$,
$b_j$ are constants. The results in the following will depend on
the ratio of these coefficients only. However, to keep the
transition to the Dirichlet and Neumann cases transparent we will
use the form (\ref{boundcond}).
   In the case of a conformally coupled
scalar the corresponding regularized VEV's for the energy-momentum
tensor are uniform in the region between the plates and have the
form
\begin{equation}\langle T_{\nu }^{\mu }\left[ \eta _{\alpha \beta }\right] \rangle _{%
{\rm ren}}=-\frac{J_1(B_1,B_2)}{2\pi ^{1/2}a^{2}\Gamma (3/2)}{\rm
diag}(1,-1), \quad r_{1}< r< r_{2}, \label{emtvevflat}
\end{equation}
where
\begin{equation}\label{IDB1B2}
  J_1(B_1,B_2)={\rm p.v.}
\int_{0}^{\infty }\frac{t dt}{\frac{(B_{1}t-1)(B_{2}t-
1)}{(B_{1}t+1)(B_{2}t+1)}e^{2t}-1},
\end{equation}
and we use the notations
\begin{equation}
B_{j}=\frac{\bar{b}_{j}}{\bar{a}_{j}a},\quad j=1,2,\quad
a=r_{2}-r_{1}. \label{Bjcoef}
\end{equation}
For the Dirichlet scalar $B_1=B_2=0$ and one has
$J_D(0,0)=2^{-2}\Gamma (2)\zeta _R(2)$, with the Riemann zeta
function $\zeta _R(s)$. Note that in the regions $r< r_{1}$ and
$r> r_{2}$ the Casimir densities vanish :
\begin{equation}
\langle \bar{T}_{\nu }^{\mu }\left[ \eta _{\alpha \beta }\right] \rangle _{%
{\rm ren}}=0,\quad r< r_{1},r> r_{2}.  \label{emtvevflat2}
\end{equation}
The two-dimensional Schwarzschild spacetime is asymptotically
flat,i.e at infinity is Minkowski spacetime, so the constants of
integration $\alpha$ and $\beta$ are evaluated demanding the
regularized stress-tensor Eq.(\ref{emteq}) to coincide with the
standard Casimir stress tensor Eq.(\ref{emtvevflat}) at infinity
$r\rightarrow \infty$. Here we introduce the state vector $|C>$,
which is the analogue of the Boulware state \cite{Bul}. In the
case of the existence of
    a boundary the Minkowski limit of $|C>$ is not the Minkowski
   state $|M>$. In this limit $<C|T_{\mu\nu}|C>$ is non-zero and shows the
    effects of the boundary conditions on the
   vacuum of the scalar filed, therefore we obtain
   \begin{equation}
  \beta=\varepsilon_{c}^{1}-\frac{1}{384\pi m^{2}},      \hspace{2cm}
  \alpha=0. \label{cof1}
  \end{equation}
  where $\varepsilon_{c}^{1}$ is given by
  \begin{equation}
\varepsilon_{c}^{1}=-\frac{J_1(B_1,B_2)}{2\pi ^{1/2}a^{2}\Gamma
(3/2)}. \label{epseq}
  \end{equation}
  Now we consider the Hartle-Hawking state $|H>$ \cite{Hart}. This state is not empty at
  infinity,even in the absence of boundary conditions on the quantum
  filed, but it corresponds to a thermal distribution of
  quanta at the Hawking temperature $T=\frac{1}{8\pi m}$. In
  fact, the state $|H>$ is related to a black hole in equilibrium
  with an infinite reservoir of black-body radiation. In the absence of
  boundary conditions the stress tensor at infinity is equal to
  \begin{equation}
  <H|T^\mu_\nu|H>=\frac{\pi T^{2}}{12}\left(\begin{array}{cc}
   2 & 0 \\
   0 &-2 \
   \end{array}\right).
  \end{equation}
 In the presence of the boundary conditions, equation (\ref{emtvevflat}) has to be
 added
  to the above
 relation. In this case at $r\rightarrow\infty$ we have
 \begin{equation}
  <H|T^{\mu}_{\nu}|H>=(\frac{1}{384\pi m^{2}}+ \varepsilon_{c}^{1})\left(\begin{array}{cc}
  1 & 0 \\
  0 & -1\
  \end{array}\right),
  \end{equation}
  then
  \begin{equation}
   \beta=\varepsilon_{c}^{1}-\frac{1}{192 \pi m^{2}},   \hspace{2cm}
   \alpha=0. \label{cof2}
  \end{equation}
 The difference between (\ref{cof2})and (\ref{cof1}) is due to the existence of the bath
  of thermal radiation at temperature
  $T$.\\
  In order to calculate the contribution from the Hawking evaporation
  process to the Casimir energy(total vacuum energy), for this special
  geometry. We introduce the final quantum state which is a
  convenient candidate for the vacuum \cite{Unruh}. This state is
  called the
  Unruh state $|U>$. In the limit $r\rightarrow \infty$, this state
  corresponds to the outgoing flux of a black-body radiation
  at black hole temperature $T$. The stress tensor in the
  limit $r\rightarrow \infty$ and in the absence of the boundary
  conditions is as follows:
  \begin{equation}
   <U|T^\mu_\nu|U>=\frac{\pi T^{2}}{12}\left(\begin{array}{cc}
   1 & 1 \\
   -1 &-1 \
   \end{array}\right).\label{unro}
  \end{equation}
  In the presence of the boundary conditions the expression (\ref{emtvevflat})
  should be added to the relation (\ref{unro}) which is a new stress tensor. Comparing
  the new stress with Eq.(\ref{emteq}) one obtains
\begin{equation}
  \alpha=\frac{-1}{768\pi m^{2}}    ,\hspace{2cm} \beta=\varepsilon_{c}^{1}
  -\frac{1}{256\pi m^{2}}  .
  \end{equation}
   Finally, the form of the stress tensor(\ref{emteq}), in, respectively Boulware ,Hartle-Hawking  and Unruh
   states is as follows:
   \bea \label{bulten}
  <B|T^{\mu}_{\nu}(r)|B>=
    \left(\begin{array}{cc}
  T^{\alpha}_{\alpha}-\Omega^{-1}(r)H(r)&0\\
  0&\Omega^{-1}H(r)\
  \end{array}\right)&& \nn\\+
   \Omega^{-1}(\frac{J_1(B_1,B_2)}{2\pi ^{1/2}a^{2}\Gamma
(3/2)}+\frac{1}{384\pi
  m^{2}})\left(\begin{array}{cc}
  1&0\\
  0&-1\
  \end{array}\right)
  \eea
  \bea \label{hawten}
  <H|T^{\mu}_{\nu}(r)|H>=\left(\begin{array}{cc}
  T^{\alpha}_{\alpha}-\Omega^{-1}(r)H(r)&0\\
  0&\Omega^{-1}H(r)\
  \end{array}\right)&& \nn\\
  +\Omega^{-1}(\frac{J_1(B_1,B_2)}{2\pi ^{1/2}a^{2}\Gamma
(3/2)}+\frac{1}{192\pi
  m^{2}})\left(\begin{array}{cc}
  1&0\\
  0&-1\
  \end{array}\right)
  \eea
  \bea \label{unten}
<U|T^{\mu}_{\nu}(r)|U>=\left(\begin{array}{cc}
  T^{\alpha}_{\alpha}-\Omega^{-1}(r)H(r)&0\\
  0&\Omega^{-1}H(r)\
  \end{array}\right)&& \nn\\+\Omega^{-1}\left(\begin{array}{cc}
  \frac{J_1(B_1,B_2)}{2\pi ^{1/2}a^{2}\Gamma
(3/2)}+\frac{1}{256\pi m^{2}}& \frac{-1}{768\pi m^{2}}\\
  \frac{1}{768\pi m^{2}}& -\frac{J_1(B_1,B_2)}{2\pi ^{1/2}a^{2}\Gamma
(3/2)}-\frac{1}{256\pi m^{2}}\
  \end{array}\right).
  \eea
  Here the presence of the form
  \begin{equation}\label{bunpar}
  \frac{J_1(B_1,B_2)}{2\pi ^{1/2}a^{2}\Gamma
(3/2)}\Omega^{-1}\left(\begin{array}{cc}
  -1&0\\
  0&1\
  \end{array}\right).
  \end{equation}
  is due to the boundary conditions. Therefore, Eqs.(\ref{bulten}-\ref{unten})are
  separable as follows:
    \begin{equation}
    <B|T^\mu  _\nu|B>=<B|  T^{(g) \mu}  _\nu|B>+<B|T^{(b)\mu}
    _\nu|B>
  \end{equation}
  \begin{equation}
    <H|T^\mu  _\nu|H>=<H|  T^{(g) \mu}  _\nu|H>+<H|T^{(b)\mu}
    _\nu|H>+<H|T^{(t)\mu}_{\nu}|H>
  \end{equation}
 \begin{equation}
    <U|T^\mu  _\nu|U>=<U|  T^{(g) \mu}  _\nu|U>+<U|T^{(b)\mu}
    _\nu|U>+<U|T^{(r)\mu}_{\nu}|U>
  \end{equation}
  where $<T^{(g) \mu}  _\nu>$,  $<T^{(b)\mu}  _\nu>$ and $<T^{(r)\mu}_{\nu}>$
  correspond to gravitational, boundary and Hawking radiation contributions, respectively,
  and  $<T^{(t)\mu}_{\nu}>$ is the
  bath of thermal radiation at temperature $T$.
    It should be noted that the trace anomaly
  has a contribution just in the first term $<T^{(g) \mu}  _\nu>$, which
   comes from the background effect not the
  boundary one. However, it has a contribution in the total Casimir
  energy-momentum tensor.
  In the regions $r <r_{1}$and $r >r_{2}$ the boundary parts are zero
  and only  the gravitational polarization parts are present.\\
  The vacuum boundary part pressures acting on plates are
  \begin{equation}\label{pereq}
  P_{b_{1,2}}=P_{b}(r=r_{1,2})=-<T^{(b)1}_{1}(r=r_{1,2})>=
  \Omega^{-1}(r_{1,2})\frac{J_1(B_1,B_2)}{2\pi ^{1/2}a^{2}\Gamma
(3/2)},
  \end{equation}
  this corresponds to the attractive/repulsive force between the plates if $%
P_{b_{1,2}}</>0$. The equilibrium points for the plates correspond
to the zero values of Eq.(\ref{pereq}): $P_{b_{1,2}}=0$. These
points are zeros of the function $J_1(B_1,B_2)$ defined by
Eq.(\ref{IDB1B2}) and are the same for both plates. The effective
pressure created by
  other parts in (\ref{bulten}-\ref{unten}) are the same for
  both sides of the plates, and hence lead to the vanishing effective force.

  \section{Conclusion}
  In the semiclassical approximation theory of quantum gravity we
are involved
 in calculating the expectation value of energy-momentum tensor in special
 vacuum \cite{Birrell}. However, the usual expression
 for the stress tensor includes singular products of the field
 operators for stress tensor. Renormalization theory of the stress
 tensor claims to solve this problem, but it must be mentioned that
 the usual scheme of renormalization includes complexity and somewhat
 ambiguity. For instance, there is no conceptual support for a local
 measure of energy-momentum of some given state without any reference
 to any global structure. We know in this frame energy is source of
 gravity and we are not allowed to subtract any unwanted part of
 energy even though it is infinite. So to consider the back-reaction effect of the
 quantum field on the gravitational field,  we must
 find a more elaborate renormalization scheme in which the dynamics of
 gravitational field is a vital component.
 In the present paper we have found the renormalized energy-momentum tensor for a massless
scalar field
  on background of two dimensional Schwarzschild black hole for two
  plates with Robin boundary conditions, by making use of general
  properties of stress tensor only.
  We propose that if we know the stress tensor for a given boundary
  in Minkowski space-time, the Casimir effect in gravitational background
  can be calculated. We have found direct relation between trace
  anomaly and total Casimir energy.
   In addition, by considering the Hawking radiation for observer
  far from black hole, this radiation
  contributes to the Casimir effect.\\
  In this paper we have derived three renormalized energy-momentum
  tensors for our case under study. This is due to selecting three
  types of
  vacuum states for our calculation. If we consider the Boulware vacuum,
   the stress tensor will have two parts: a boundary part and a
  gravitational part. However, using Hartle-Hawking  and Unruh vacuums will result
  in another term being added to the stress tensor, which, respectively, corresponds to
  a bath of thermal radiation and Hawking radiation. In the region
between the plates the boundary induced part for the vacuum
energy-momentum tensor is given by Eq.(\ref{bunpar}), and the
corresponding  vacuum forces acting on the plates have the form
Eq.(\ref{pereq}). These forces vanish at the zeros of the
function $J_1(B_1,B_2)$. For a conformally coupled massless
scalar field with Robin boundary condition this effect was
initially studied in Ref.\cite{shse} for a background
Randall--Sundrum geometry \cite{Rand99}. Therefore in this case
the Casimir effect provide a  possibility for the stabilization of
the distance (radion field) between the branes (for more study
see \cite{eliz1}).
 The effective pressure created by
  other parts in Eqs.(\ref{bulten}, \ref{hawten}, \ref{unten}) are the same from the
  both sides on the plates, and hence leads to the zero effective force.

\end{document}